\documentclass[journal]{IEEEtran}
\IEEEoverridecommandlockouts
\IEEEaftertitletext{\vspace{-2.0\baselineskip}}
\usepackage{cite}
\usepackage{amsmath,amssymb,amsfonts}
\usepackage{algpseudocode}
\usepackage{graphicx}
\usepackage{textcomp}
\usepackage{algorithm}
\usepackage{algpseudocode}
\usepackage{xcolor}
\usepackage{subcaption} 

\def\BibTeX{{\rm B\kern-.05em{\sc i\kern-.025em b}\kern-.08em
    T\kern-.1667em\lower.7ex\hbox{E}\kern-.125emX}}

\usepackage[font=footnotesize]{caption}
\usepackage{titlesec}
\titlespacing{\section}{0pt}{.5ex plus .2ex minus .2ex}{0.5ex}
\titlespacing{\subsection}{0pt}{.4ex plus .2ex minus .2ex}{0.4ex}
\usepackage{mathtools}

\begin{document}

\title{Channel Knowledge Map Enabled Low-Altitude ISAC Networks: Joint Air Corridor Planning and Base Station Deployment\\
\author{Jiaxuan Li,
        Yilong Chen,
        Fan Liu,~\IEEEmembership{Senior~Member,~IEEE},
        and~Jie~Xu,~\IEEEmembership{Fellow,~IEEE}}
\thanks{J. Li, Y. Chen, and J. Xu are with the School of Science and
 Engineering (SSE), the Shenzhen Future Network of Intelligence Institute
 (FNii-Shenzhen), and the Guangdong Provincial Key Laboratory of
 Future Networks of Intelligence, The Chinese University of Hong Kong
 (Shenzhen), Guangdong, 518172, China (e-mail: jiaxuanli4@link.cuhk.edu.cn; yilongchen@link.cuhk.edu.cn;
 xujie@cuhk.edu.cn).
 
F. Liu is with the National Mobile Communications Research Laboratory, School of Information Science and Engineering, Southeast University, Nanjing 210096, China (e-mail: f.liu@ieee.org).

F. Liu is the corresponding author.}
}

\maketitle
%\iffalse
\begin{abstract}
This letter addresses the joint air corridor planning and base station (BS) deployment problem for low-altitude integrated sensing and communication (ISAC) networks. In the considered system, unmanned aerial vehicles (UAVs) operate within a structured air corridor composed of connected cubic segments, and multiple BSs need to be selectively deployed at a set of candidate locations to ensure both sensing and communication coverage throughout the corridor. In particular, we leverage the channel knowledge map (CKM) to characterize wireless channels for candidate BS sites prior to deployment, thereby facilitating the offline planning. Under this setup, we minimize the system cost in terms of the weighted sum of the air corridor length and the number of deployed BSs, subject to the constraints on both sensing and communication performance across the corridor. To solve the formulated large-scale nonconvex integer programming problem, we develop a hierarchical coarse-to-fine grid decomposition algorithm. Simulation results demonstrate the benefit of the proposed joint design in reducing the overall deployment cost while ensuring the coverage of the low-altitude ISAC networks.
\end{abstract}


\section{Introduction}

Low-altitude economy (LAE) leverages unmanned aerial vehicles (UAVs) and electric vertical takeoff and landing (eVTOL) aircraft for urban air mobility and cargo delivery. Air corridors and integrated sensing and communication (ISAC) have emerged as two promising techniques to ensure the safe and reliable operations of UAVs and eVTOLs. On the one hand, air corridors provide structured routing to deconflict traffic and simplify airspace management \cite{mobility}. On the other hand, ISAC networks enable simultaneous sensing and communication within air corridors, supporting real-time tracking, navigation, and reliable control of UAVs and eVTOLs \cite{Integrated_Sensing_and_Communications}. As such, the integration of air corridors and ISAC is crucial for scalable and safe LAE.

In the literature, there have been various prior works investigating air corridors \cite{aircorridor,coordinate} and ISAC networks \cite{overview1,overview2,Cooperative_ISAC_Networks} for LAE separately. First, the authors in \cite{aircorridor,coordinate} studied the design of geometric configurations (e.g., cubic or cylindrical segments) of air corridors and the corresponding routing strategies, with the objective of maximizing the traffic throughput while ensuring collision-free operations. Next, various prior works (e.g., \cite{overview1,overview2,Cooperative_ISAC_Networks}) exploited on-ground ISAC networks to support low-altitude UAV operations via ground-based cellular infrastructure. For example, the authors in \cite{Cooperative_ISAC_Networks} utilized the cooperative transmit beamforming at multiple on-ground base stations (BSs) to support the communication and localization of cooperative UAVs and monitor the low-altitude airspace to prevent the invasion of unauthorized flight objects. Despite these advances, the joint optimization of air corridor planning and ISAC network deployment remains an open problem, which motivates this work.

Specifically, we study joint design of air corridors and low-altitude ISAC networks, for providing seamless sensing and communication coverage throughout air corridors. In this context, obtaining channel state information (CSI) with potential BS sites is particularly challenging due to the lack of prior measurements. Channel knowledge map (CKM) has emerged as a promising solution, which establishes a functional mapping from transceiver locations to the corresponding channel knowledge. In practice, CKM can be effectively constructed based on physical environment knowledge and limited channel measurements \cite{b5,CKM1,CKM2}, providing CSI {\it a priori} to facilitate network deployment and operation\cite{b4,CKM}. However, how to exploit CKM to facilitate the joint optimization of air corridor planning and BS deployment has not been studied in the literature yet.

This letter studies the CKM-enabled joint air corridor planning and BS deployment problem for low-altitude ISAC networks at the planning stage, providing a stable upper-layer infrastructure and corridor layout that can be further adapted in operation to dynamic urban environments. To be specific, we model the air corridor as a sequence of connected cubes with a fixed edge length located at a fixed altitude, and we need to selectively deploy multiple BSs at a set of candidate locations to provide communication and sensing coverage across the corridor. For each candidate BS site, we utilize the CKM to predict the CSI and evaluate the communication and sensing coverage performance, thereby enabling offline optimization. Under this setup, we aim to minimize the system cost in terms of the weighted sum of the air corridor length and the number of deployed BSs, subject to the sensing and communication coverage requirements throughout the corridor. We propose a hierarchical coarse-to-fine grid decomposition algorithm to solve the formulated large-scale nonconvex integer programming problem. In the coarse layer, we reformulate the problem as a linear integer program to obtain a feasible air corridor and BS deployment solution. In the fine layer, we use an alternating optimization (AO) procedure to refine the coarse layer solution for further reducing the system cost. Simulation results demonstrate that the proposed algorithm achieves shorter air corridor lengths and requires fewer BSs to fulfill the LAE tasks, thereby significantly reducing system costs compared with various baseline methods.

\section{System Model and Problem Formulation}
\begin{figure}[!htbp]
    \centering
    \includegraphics[width=0.75\linewidth]{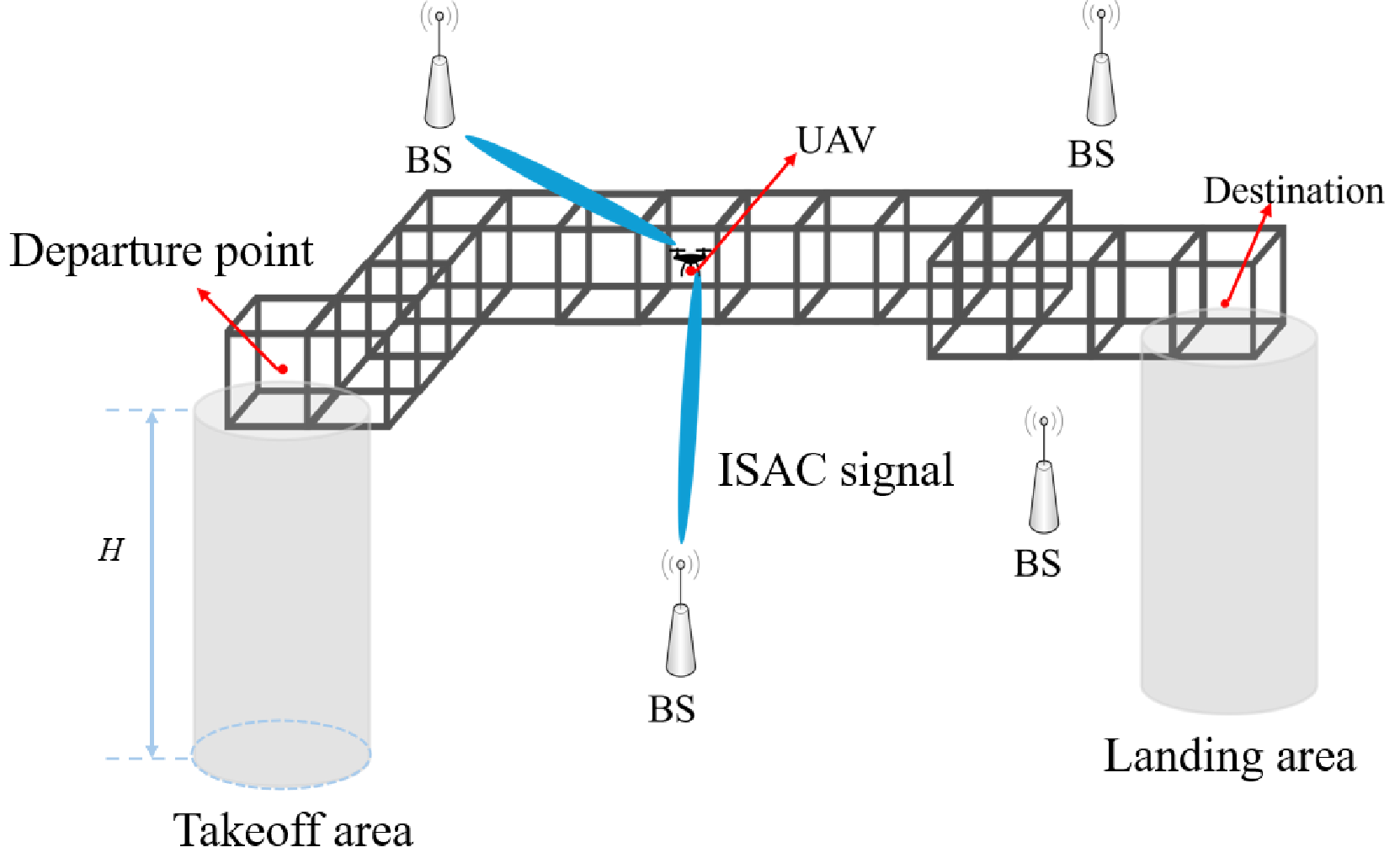}
    %\captionsetup{justification=raggedright, singlelinecheck=false}
    \caption{System model.}
    \label{fig:enter-label}
\end{figure}

We consider an ISAC network for LAE as illustrated in Fig. 1, in which potential UAVs pass through an urban area from a departure point to a destination point within a dedicated air corridor (to be designed) to fulfill various tasks such as urban air mobility and cargo delivery. To ensure reliable communication and real-time sensing for UAVs, multiple ground BSs are selectively deployed at a set of candidate locations on the ground to support these services throughout the corridor.

\subsection{Air Corridor Model}

An air corridor is a three-dimensional (3D) volume of airspace reserved for UAV and eVTOL operations \cite{aircorridor}. In this work, we consider a corridor at a fixed-altitude $H$.\footnote{The considered system can also be viewed as one representative air corridor among multiple corridors in a multi-layer air corridor architecture.} The vertical takeoff and landing zones are assumed to lie directly beneath the departure and destination points. Under these conditions, our objective is to design an air corridor that safely accommodates flights within the prescribed airspace.

As shown in Fig. 1, we adopt a discretized grid map with Cartesian coordinates, and the air corridor is modeled as a sequence of uniformly sized cubes connected face to face to form a continuous path \cite{aircorridor}. The grid consists of an \(N\times N\) 3D rectangular region at a constant altitude \(H\), indexed by \(\{\mathcal{R}_{i,j}\mid i,j\in \mathcal{N}\},\ \mathcal{N}\triangleq \{1,2,\dots,N\}\), with the origin at $(x^{O},y^{O},H)$. The geographical coverage of each grid cell \(\mathcal{R}_{i,j}\) is defined as
\begin{equation}
\begin{aligned}
\mathcal R_{i,j}
=&
\bigl[x^{O} + (i-1)\,\Delta_x,\;x^{O} + i\,\Delta_x\bigr]\\
\;\times\;
&\bigl[y^{O} + (j-1)\,\Delta_y,\;y^{O} + j\,\Delta_y\bigr]
\;\times\;
\,[H,H+\Delta_z],
\end{aligned}
\end{equation}
where $\Delta_x, \Delta_y$, and $\Delta_z$ denote the grid resolutions along the $x$-, $y$-, and $z$-axes, respectively.

Furthermore, let $\mathbf{A}$ denote an \(N \times N\) binary indicator matrix, with \(A_{i,j}=1\) if \(\mathcal{R}_{i,j}\) is included in the air corridor, and \(A_{i,j}=0\) otherwise. Without loss of generality, we designate $\mathcal{R}_{1,1}$ and $\mathcal{R}_{N,N}$ as the departure and destination points of the corridor, respectively. We thus have
\begin{equation}
    A_{1,1}=1,\quad A_{N,N}=1. \label{connec3}
\end{equation}

Note that in order for the air corridor to be feasible, a necessary condition is that the corridor should  form a continuous path, meaning that all active cells with $A_{i,j}=1$ must be connected. This is equivalent to the condition that each active cell $(i,j)$ with\ $A_{i,j}=1$ should have at least two active adjacent grid along the $x$- or $y$-axis. Moreover, to avoid unnecessary branching, each active cell is restricted to have at most two active neighbors. To this end, for each grid \((i,j)\), we define the set of four neighboring grid cells as $\mathcal{N}_{4}(i,j)\triangleq\{(i',j')\in\mathcal{N}\times\mathcal{N}:\ |i'-i|+|j'-j|=1\}$. Accordingly, the connectivity constraints are expressed as
\begin{equation}
    2 A_{i,j}\leq\sum\nolimits_{(i',j')\in \mathcal N_4(i,j)}A_{i',j'} \leq2.\label{connec1}
\end{equation}

Additionally, to guarantee reachability, there must exist at least one active grid in each row and column between the departure point and the destination point. We have
\begin{equation}
\begin{aligned}
    \sum\nolimits_{i=1}^{N} A_{i,j} \ge 1,\quad \forall j\in\mathcal N,\\
    \sum\nolimits_{j=1}^{N} A_{i,j} \ge 1,\quad \forall i\in\mathcal N.\label{connec4}
\end{aligned}     
\end{equation}

\subsection{Sensing and Communication Models}

Suppose that there are \(K\) candidate BS sites with indices \(\mathcal{K}\triangleq\{1,\cdots,K\}\). Let \(\mathbf{q}_k=(q_{k,x},q_{k,y},q_{k,z})\) denote the location of site \(k,k\in\mathcal{K}\). We introduce a binary vector \(\boldsymbol{\delta}=(\delta_1,\cdots,\delta_K)\) to indicate the deployment status of candidate sites, where $\delta_k = 1$ if a BS is deployed at candidate site \(k\) and $\delta_k = 0$ otherwise. 

To characterize the sensing and communication coverage throughout the air corridor, we need to acquire CSI from each candidate BS site to any point within the air corridor. However, acquiring such measurements at a dense set of locations is typically impractical, especially when the BS has not been properly deployed. To address this, we exploit the CKM to obtain channel power gains. In practice, we can first obtain the 3D environmental model of the considered region, and then utilize the ray tracing technique to construct the channel gain map \cite{sionna}. \footnote{The accuracy of ray tracing process highly depends on the accuracy of the 3D environment model. In order to focus our study on the network deployment design, we assume that the CKM is perfectly known a {\it priori} and leave the CKM construction and the analysis of inaccurate CKM for future work.} It is assumed that for each candidate BS site, we use the CKM to obtain the mapping from a 3D user location \(\{\mathbf{p}=(q_x,q_y,q_z)\}\) to the corresponding channel power gain at that BS site. Let \(h_k(\mathbf{p})\) denote the channel power gain between BS site $k\in\mathcal{K}$ and location \(\mathbf{p}\), which in general is a discrete function without a closed-form expression. In particular, for the rectangular region $\mathcal{R}_{i,j}$ associated with $A_{i,j}$, \(i,j\in\mathcal{N}\), the set of sampled channel power gains from BS site $k$, $k\in\mathcal{K}$ within $\mathcal{R}_{i,j}$ is 
\begin{equation}
    \mathcal{G}_{k,i,j} = \;\bigl\{\,h_k(\mathbf{p})\mid \mathbf{p}\in\mathcal{R}_{i,j}\bigr\}.
\end{equation}

For communication, by leveraging the CKM, we evaluate the received signal-to-interference-plus-noise ratio (SINR) at sample positions \(\mathbf{p}\) as the communication performance metric. By considering that the receiver is deployed with a single antenna, the SINR of the arrival signal at location \(\mathbf{p}\) from BS \(k\) is given by
\begin{equation}
    \gamma_{k,i,j}(\mathbf{p},\boldsymbol{\delta})
=\frac{P\,\delta_{k}Gh_{k}(\mathbf{p})}
      {\sum_{b\in\mathcal{K}\setminus k}\delta_{b}\,Ph_{b}(\mathbf{p})+\sigma^{2}},
\end{equation}
where $P$ is the transmit power of each BS, $G>1$ is a coefficient capturing the transmit antenna gain and the potential transmit beamforming gain, and \(\sigma^2\) is the noise power at the receiver. To ensure the coverage within each grid cell \((i,j)\), we take the worst-case SINR over all sampling points within this cell as the communication performance metric. Specifically, for all points within $\mathcal{R}_{i,j}$, we define \(h_{k,i,j}^{\min}\triangleq\min\;(\mathcal{G}_{k,i,j})\) and \(h_{k,i,j}^{\max}\triangleq\max\;(\mathcal{G}_{k,i,j})\) as the minimum and maximum channel power gains from BS site \(k\), respectively. Accordingly, the worst-case SINR is given by
\begin{equation}
    \underline{\gamma}_{k,i,j}(\boldsymbol{\delta})\triangleq\frac{PG\delta_kh_{k,i,j}^{\min}}{\sum_{b\in\mathcal{K}\setminus k}\delta_bPh_{b,i,j}^{\max}+\sigma^{2}}.\label{SINR}
\end{equation}

In addition, we consider the line-of-sight (LoS) sensing to localize aircraft within the air corridor for navigation and surveillance. To this end, at least three LoS paths are needed, and the total received echo signal power from LoS paths should exceed a certain threshold \cite{loc}. We define an indicator \(\mathcal{L}_{k,i,j}\), where \(\mathcal{L}_{k,i,j}=1\) if there exists a LoS link between BS site \(k\) and location \(\mathbf{p}\in\mathcal{R}_{i,j}\), and \(\mathcal{L}_{k,i,j}=0\) otherwise. Assuming a monostatic LoS sensing model, we consider the minimum echo signal power over \(\mathcal{R}_{i,j}\) from BS site \(k\) as the sensing performance metric, which is given by \cite{radar}
\begin{equation}
p_{k,i,j}\triangleq\min_{\mathbf{p}\in\mathcal{R}_{i,j}}\frac{\mathcal{L}_{k,i,j}PG\lambda^2\sigma_{RCS}}{(4\pi)^3d_{k}^{4}(\mathbf{p})},\label{power}
\end{equation}
where $d_{k}(\mathbf{p})\triangleq\Vert \mathbf{q}_k-\mathbf{p}\Vert_2$ is the distance between BS site \(k\) and location \(\mathbf{p}\in\mathcal{R}_{i,j}\), \(\lambda\) is the wavelength, and \(\sigma_{\text{RCS}}\) denotes the radar cross section of potential UAV.

\subsection{Problem Formulation}

In this work, we aim to minimize the system cost in terms of the weighted sum of the air corridor length and the number of deployed BSs, subject to the path connectivity, SINR, LoS visibility, and sensing power constraints. The problem is formulated as
\begin{subequations}
\begin{align}
(\text{P1}):\quad&\min_{\substack{\mathbf{A},\boldsymbol{\delta}}}\quad  \alpha_1 \mathbf{1}^{\top}\mathbf{A}\mathbf{1} + \alpha_2 \sum\nolimits_{k=1}^K \delta_k, \notag\\
    \text{s.t.}\;\;
    &\sum\nolimits_{k=1}^{K} \delta_k p_{k,i,j} \geq \epsilon_1 A_{i,j}, \ \forall i,j\in\mathcal{N},\label{power1}\\
    &\sum\nolimits_{k=1}^{K} \delta_k \mathcal{L}_{k,i,j} \geq 3 A_{i,j}, 
    \ \forall i,j\in\mathcal{N},\label{los1}\\
    &\max\nolimits_{k\in\mathcal{K}} \underline{\gamma}_{k,i,j}(\boldsymbol{\delta})\geq \epsilon_2 A_{i,j}, 
    \ \forall i,j\in\mathcal{N}\label{sinr_o},\\
    &\delta_k \in \{0, 1\}, A_{i,j} \in \{0, 1\}, 
    \ \forall k\in\mathcal{K}, i,j\in\mathcal{N}\label{integ1},\\
    &\quad \eqref{connec3}-\eqref{connec4},\notag
\end{align}
\end{subequations}
where $\alpha_1\ge 0$ and $\alpha_2 \geq 0$ are weighting coefficients determined by the economics and policy constraints of the target low-altitude network with $\alpha_1+\alpha_2=1$, and $\mathbf{1}\in\mathbb{R}^N$ denotes the all-ones vector, with $\mathbf{1}^{\top}$ being its transpose. For any corridor cell $(i,j)$ (i.e., $A_{i,j}=1$), the constraints in~\eqref{power1} and \eqref{sinr_o} ensure received sensing power and communication SINR no less than the corresponding thresholds $\epsilon_1$ and $\epsilon_2$, respectively, while the constraint in \eqref{los1} guarantees the visibility from at least three LoS BSs.

Problem (P1) is an integer nonconvex program with hard integrality constraints that is inherently difficult to solve using relaxation-based methods. Note that the decision space consists of $N^{2}+K$ binary variables. Consequently, in practical settings with fine spatial resolution, (P1) becomes a large-scale problem, significantly increasing computational complexity.

\section{Proposed Solution to Problem (P1)}
In this section, we eliminate the nonconvex constraint in \eqref{sinr_o}, yielding a linear integer program. Despite this reformulation, the problem still remains computationally demanding due to its scale and integrality. To address this, we adopt a coarse-to-fine hierarchical framework that first obtains a jointly optimized coarse-resolution path and BS deployment, and then refines the solution at the fine layer to improve spatial coverage with reduced BS deployment overhead.

\subsection{Joint Optimization over Coarse Grids} 

At the coarse layer, the original $N\times N$ map $\mathbf A$ is coarsened to an $M\times M$ grid $\mathbf{B}$, with index set $\mathcal M\triangleq\{1,\ldots,M\}$. Each coarse cell $B_{a,b}$, $a,b\in\mathcal M$ represents a rectangular region $\tilde{\mathcal {R}}_{a,b}$ with spatial resolutions $\tilde{\Delta}_x, \tilde{\Delta}_y,$ and $\Delta_z$. Let $\boldsymbol{\delta}^{B}$ denote the associated BS deployment vector. For each $(a,b)$, we evaluate the cell SINR $\underline{\tilde{\gamma}}_{k,a,b}(\boldsymbol{\delta}^{B})$ according to \eqref{SINR}, by using the minimum and maximum channel power gains \(\tilde {h}_{k,a,b}^{\min}\) and \(\tilde{h}_{k,a,b}^{\max}\) based on the sample set \(\tilde{\mathcal{G}}_{k,a,b}\) within $\tilde{\mathcal {R}}_{a,b}$. To suppress outliers in each coarse cell, we discard the lowest 10\(\%\) and the highest 10\(\%\) of the samples in \(\tilde{\mathcal{G}}_{k,a,b}\) and compute \(\tilde {h}_{k,a,b}^{\min}\) and \(\tilde{h}_{k,a,b}^{\max}\) on the remainder. We also define \(\tilde{\mathcal{L}}_{k,a,b}\) and \(\tilde{p}_{k,a,b}\) in this layer, following the same form as in Section II-B.

We address the joint optimization problem (P1) at the re-partitioned coarse layer, with the newly defined indicator matrix \(\mathbf{B}\) and deployment indicator vector \(\boldsymbol{\delta}^{B}\). Firstly, to eliminate the max operator in \eqref{sinr_o}, we apply a Big-$M$ reformulation~\cite{bigM}. Specifically, we introduce binary auxiliary variables $\{z_{k,a,b}\in\{0,1\}| a, b \in\mathcal{M}, k\in\mathcal{K}\}$ to indicate whether BS site $k$ satisfies the SINR threshold at cell $(a,b)$. Then, for each coarse cell \((a,b),a,b \in \mathcal{M}\), the constraint \(\max_{k\in\mathcal{K}} \tilde{\underline{\gamma}}_{k,a,b}(\boldsymbol{\delta}^{B})\geq \epsilon_2 B_{a,b} \) is equivalently enforced by
\begin{subequations}\label{eq:bigM}
\begin{align}
\,(1-z_{k,a,b})\zeta
&\;\ge\;
\epsilon_2\,B_{a,b}(P\!\!\sum\nolimits_{k'\in\mathcal{K}\setminus k}\!\delta_{k'}^{B}\,\tilde{h}_{k',a,b}^{\max}+\sigma^2)\notag\\
&-P\,G\,\delta_k^{B}\,\tilde{h}_{k,a,b}^{\min} , \label{eq:bigM_link}\\[2pt]
z_{k,a,b} &\le \delta_k^{B},  \label{eq:bigM_link2}\\[2pt]
\sum\nolimits_{k=1}^{K} z_{k,a,b} &\ge B_{a,b},\label{eq:bigM_any}\\
\qquad z_{k,a,b}&\in\{0,1\}.
\end{align}
\end{subequations}
Here, $\zeta>0$ is a sufficiently large constant (chosen larger than an upper bound on the right-hand side of (10a)) used to deactivate the SINR constraint in \eqref{eq:bigM_link} when $z_{k,a,b}=0$, and to activate the constraint when $z_{k,a,b}=1$. \eqref{eq:bigM_link2} prevents the selection of an undeployed BS, and \eqref{eq:bigM_any} ensures that at least one deployed BS satisfies the threshold whenever $B_{a,b}=1$.

The constraint \eqref{eq:bigM_link} remains nonconvex because it contains the bilinear products $B_{a,b}\,\delta_{k'}^{B}$. Since all these variables are binary, we introduce auxiliary binaries $w_{k,a,b}=B_{a,b}\,\delta_{k}^{B}$ for all $k\in\mathcal{K}$ and $a,b\in\mathcal{M}$. We enforce this identity $w_{k,a,b}=B_{a,b}\,\delta_{k}^{B}$ through the standard McCormick linearization \cite{four}:
\begin{subequations}\label{eq:fortet}
\begin{align}
w_{k,a,b} &\le \delta_{k}^{B},\\
w_{k,a,b} &\le B_{a,b},\\
w_{k,a,b} &\ge \delta_{k}^{B} + B_{a,b} - 1,\\
w_{k,a,b} &\in \{0,1\}. \label{four_linear}
\end{align}
\end{subequations}
This linearization guarantees that \(w_{k,a,b} = 1\) if and only if both \(B_{a,b} = 1\) and \(\delta_k^{B} = 1\), and \(w_{k,a,b} = 0\) otherwise. By substituting \(w_{k',a,b}\) for \(B_{a,b}\delta_{k'}^{B}\), \eqref{eq:bigM_link} is rewritten as
% \begin{align}
% P\,G\,\delta_k^{B}\,\tilde{h}_{k,a,b}^{\min}
% &\;\ge\;
% \epsilon_2(
% P\sum_{k'\in\mathcal{K}\setminus k} \tilde{h}_{k',a,b}^{\max}\,w_{k',a,b}
% + \sigma^2 B_{a,b}) \notag\\
% &- M(1 - z_{k,a,b}),\forall\,k\in\mathcal{K},a,b\in\mathcal{M}. \label{eq:bigM_link_lin}
% \end{align}

\begin{align}
(1 - z_{k,a,b})\zeta
&\;\ge\;
\epsilon_2(
P\sum\nolimits_{k'\in\mathcal{K}\setminus k} \tilde{h}_{k',a,b}^{\max}\,w_{k',a,b}
+ \sigma^2 B_{a,b}) \notag\\
&-P\,G\,\delta_k^{B}\,\tilde{h}_{k,a,b}^{\min} ,\forall\,k\in\mathcal{K},a,b\in\mathcal{M}. \label{eq:bigM_link_lin}
\end{align}
After that, at coarse layer, problem (P1) is reformulated as
\begin{subequations}
\begin{align}
    &(\text{P2}):\min_{\substack{\mathbf{B},\boldsymbol{\delta}^{B}\\\{z_{k,a,b},\forall k\in\mathcal{K}, a,b\in\mathcal{M}\}\\\{w_{k,a,b},\forall k\in\mathcal{K}, a,b\in\mathcal{M}\}}}  \alpha_1 \mathbf{1}^{\top}\mathbf{B}\mathbf{1} + \alpha_2 \sum\nolimits_{k=1}^K \delta_k^{B},\notag\\
    &\text{s.t.}\ 2 B_{a,b}\leq\sum\nolimits_{(a',b')\in \mathcal N_4(a,b)}B_{a'b'} \leq2,\forall a,b\in\mathcal{M}\\
    &\sum\nolimits_{a=1}^{M}B_{a,b}\ge1,\forall b\in\mathcal{M},\sum\nolimits_{b=1}^{M}B_{a,b}\ge1,\forall a\in\mathcal{M}\\
    &B_{1,1}=1,\quad B_{M,M}=1,\\
    &\sum\nolimits_{k=1}^{K} \delta_k^{B} \tilde{p}_{k,a,b} \geq \epsilon_1 B_{a,b}, \ \forall a,b\in\mathcal{M},\\
    &\sum\nolimits_{k=1}^{K} \delta_k^{B} \tilde{\mathcal{L}}_{k,a,b} \geq 3 B_{a,b}, 
    \ \forall a,b\in\mathcal{M},\\
    &\delta_k^{B} \in \{0, 1\}, B_{a,b} \in \{0, 1\}, 
    \ \forall k\in\mathcal{K}, a,b\in\mathcal{M}\\
    &\eqref{eq:bigM_link2}-\eqref{four_linear},\eqref{eq:bigM_link_lin}.\notag
\end{align}
\end{subequations}

The coarse layer formulation (P2) reduces the joint optimization problem to a linear integer program with $K M^{2}$ variables, compared to $K N^{2}$ in a direct fine layer formulation of (P1), where \(N\gg M\). %The resulting solution of (P2) will be refined in the fine layer next.

\subsection{Alternative Refinement of Air Corridor and BS Deployment} 

In the coarse layer, $B_{a,b}=1$ indicates that the air corridor passes through grid \((a,b)\). Let $\mathcal{B} = \bigl\{(a,b)\mid B_{a,b}=1\bigr\}$ denote the set of selected coarse-grid cells obtained by solving (P2). For each $(a,b)\in\mathcal B$, we define $\mathbf A^{(a,b)}$ as the $(N/M)\times (N/M)$ (assuming \(N\) is divisible by \(M\)) submatrix of $\mathbf A$ for the region covered by $B_{a,b}$, where each entry $A^{(a,b)}_{i,j},i,j\in\mathcal{N}^{(a,b)},$ provides the unit-cube resolution for final planning. We optimize the path within each $\mathbf{A}^{(a,b)}$, $(a,b)\in \mathcal{B}$ and update the global BS deployment $\boldsymbol\delta$ based on $\mathbf{A}$, which is assembled from all $\mathbf{A}^{(a,b)}$ via an AO framework. 

First, we initialize the fine layer deployment vector $\boldsymbol\delta$ as the coarse layer solution $\boldsymbol\delta^{B}$. Next, for each $\mathbf{A}^{(a,b)}$, $(a,b)\in \mathcal{B}$, we predefine its start and end points based on its adjacent coarse-grid neighbors. For example, if $B_{a,b}=1,\  B_{a-1,b}=1,$ and $B_{a,b+1}=1$, then cell \((a,b)\) is connected to both its southern and eastern neighbors. We then set the midpoint of $\mathbf A^{(a,b)}$'s southern edge as the start point \(A^{(a,b)}_{sx,sy}\), and the midpoint of its eastern edge as the destination point\(A^{(a,b)}_{dx,dy}\). Fixing $\boldsymbol\delta$ and these end points, we solve the following subproblem (P3.1) independently for each $\mathbf{A}^{(a,b)},(a,b)\in\mathcal{B}$.

    \begin{align}
    &(\text{P3.1}):\quad\min_{\substack{\mathbf{A}^{(a,b)}\\}}  \quad\mathbf{1}^{\top}\mathbf{A}^{(a,b)}\mathbf{1},\notag\\
        &\text{s.t.}\ 2A_{i,j}^{(a,b)}
   \le \!\!\!\sum_{\mathclap{(i',j')\in \mathcal N_4(i,j)\cap\mathcal N^{(a,b)}}}\!\!\!
   A_{i',j'}^{(a,b)} \le 2, \forall i,j\in\mathcal{N}^{(a,b)},\\
        &\sum_{i=sx}^{dx} A^{(a,b)}_{i,j} \ge 1,\forall j\in\mathcal N^{(a,b)},
\sum_{j=sy}^{dy} A^{(a,b)}_{i,j} \ge 1,\forall i\in\mathcal N^{(a,b)},\\
        &A^{(a,b)}_{sx,sy}=1,\quad A^{(a,b)}_{dx,dy}=1,\\
        &\sum\nolimits_{k=1}^{K} \delta_k p^{(a,b)}_{k,i,j} \geq \epsilon_1 A^{(a,b)}_{i,j}, \forall i,j\in\mathcal{N}^{(a,b)},\\
        &\sum\nolimits_{k=1}^{K} \delta_k \mathcal{L}^{(a,b)}_{k,i,j} \geq 3 A_{i,j}^{(a,b)}, \quad\forall i,j\in\mathcal{N}^{(a,b)},\\
        &\max\nolimits_{k\in\mathcal{K}}\  \underline{\gamma}^{(a,b)}_{k,i,j}\geq \epsilon_2 A_{i,j}^{(a,b)}, \forall i,j\in\mathcal{N}^{(a,b)},\\
        &A_{i,j}^{(a,b)}\in\{0,1\},\forall i,j\in\mathcal{N}^{(a,b)},
    \end{align}
where $p^{(a,b)}_{k,i,j}$, $\mathcal{L}^{(a,b)}_{k,i,j}$, and $\underline{\gamma}^{(a,b)}_{k,i,j}$ follow the definitions in Section II-B and are evaluated within \(\tilde{\mathcal{R}}_{a,b}\) at the fine layer resolution. Note that \(A_{1,1}^{(1,1)}\) and \(A_{N/M,N/M}^{(M,M)}\) have only one active adjacent neighbor. 

Once all $\mathbf{A}^{(a,b)}, (a,b)\in\mathcal{B}$ have been updated, the global fine layer map $\mathbf{A}$ is reconstructed by assembling the submatrices. Subsequently, we solve problem (P3.2) to update $\boldsymbol\delta$ based on the obtained $\mathbf{A}$. 
\begin{subequations}
\begin{align}
&(\text{P3.2}):\quad\quad\quad\min_{\boldsymbol{\delta}}\sum\nolimits_{k=1}^{K}\delta_k,\notag\\
&\sum\nolimits_{k=1}^{K} \delta_k p_{k,i,j} \geq\epsilon_1 A_{i,j}, \quad \forall i,j\in{\mathcal{N}},\\
& \sum\nolimits_{k=1}^{K} \delta_k \mathcal{L}_{k,i,j} \geq 3 A_{i,j}, \quad \forall i,j\in{\mathcal{N}},\\
&\max\nolimits_{k \in\mathcal{K}} \underline{\gamma}_{k,i,j}\geq \epsilon_2 A_{i,j}, \quad \forall i,j\in{\mathcal{N}},\\
&\delta_k\in\{0,1\}, \forall k\in\mathcal{K}.
\end{align}
\end{subequations}

Both (P3.1) and (P3.2) are linear integer programs. Since each AO step leads to monotonically non-increasing system cost (i.e., the air corridor length or the number of deployed BSs), and both objectives are bounded below, the algorithm is guaranteed to converge~\cite{convex}.

%Algorithm 1 summarizes the workflow for solving (P1). 

\iffalse
\begin{algorithm}
  \caption{Solving Problem (P1)}\label{alg:cap}
  \begin{algorithmic}[1]
    \State Solve the problem (P2) using the coarse layer grid CKM to get the air corridor $\mathbf{B}$ and initial BS deployment $\boldsymbol\delta^B$. Set convergence threshold $\epsilon_1,\epsilon_2 > 0$, maximum iteration number \(I\). 
    \State Set iteration index \(t=1\).
    \Repeat
      \State Update \((\mathbf{C}^{(a,b)})^{(t)}\) by solving problem (P3.1) based on \((\boldsymbol{\delta}^{C})^{(t-1)}\).
      \State Update \((\boldsymbol{\delta}^{C})^{(t)}\) by solving problem (P3.2) based on \((\mathbf{C}^{(a,b)})^{(t)}\).
      \State Update \(t=t+1\)
    \Until{$\|(\boldsymbol{\delta}^{C})^{(t)}-(\boldsymbol{\delta}^{C})^{(t-1)}\|\le\epsilon_1,\quad \|\mathbf{C}^{(a,b)})^{(t)}-(\mathbf{C}^{(a,b)})^{(t-1)}\|\le\epsilon_2,\ \forall (a,b)\in\mathcal{S}$, or \(t=I\) }
    \State \textbf{return} \ $\mathbf{C}^{(a,b)},(a,b)\in\mathcal{S}$, $\quad\boldsymbol\delta^C$
    
  \end{algorithmic}
\end{algorithm}
\fi
\section{Numerical Results}

This section presents numerical results to evaluate the performance of our proposed algorithms. We consider a \(500\) m\(\times\)500 m area in Futian District, Shenzhen, China, with $K = 30$ candidate BSs, as shown in Fig. 2. We use the open-source library Sionna RT (v0.18.0) to generate CKMs \cite{sionna}. The carrier frequency is set to be 1 GHz, the noise power at the UAV receiver is \(\sigma^2=-110\) dBm, and the transmit power of each BS is \(P=30\) dBm. We set the transmit gain of each BS to $G=12$ dB and the radar cross section to $\sigma_{\text{RCS}} = 1\ \text{m}^2$.
The altitude of air corridor is set as \(H=150\) m, and the BS height is set as 25 m. 
\begin{figure}
    \centering
    \includegraphics[width=0.75\linewidth]{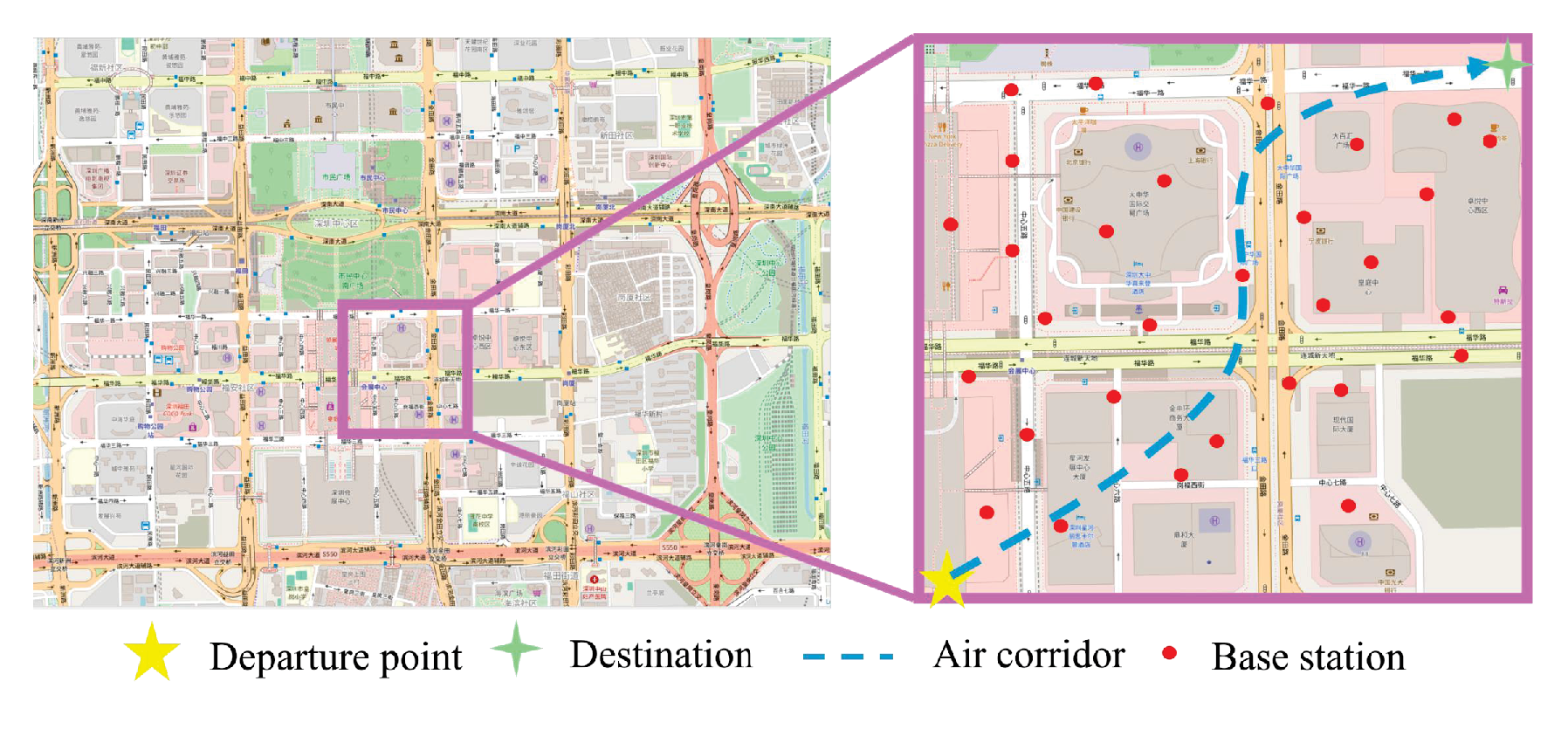}
    %\captionsetup{justification=raggedright, singlelinecheck=false}
    \caption{The considered area (from the OpenStreetMap database) and the spatial distribution of candidate BS sites.}
    \label{fig:enter-label}
\end{figure}

In the coarse layer, the map is divided into \(10\times10\) grids, with each grid measuring \(50\) m \(\times\) \(50\) m. For problem (P2), we set the weights as \(\alpha_1=0.5\) and \(\alpha_2=0.5\), and choose the Big-M constant \(\zeta=10^{-8}\). In the fine layer, the map is divided into \(100\times100\) grids, each with resolution \(\Delta_x=\Delta_y=\Delta_z=5\) m, which is consistent with the size of the air corridor cube.\footnote{We target small-UAV corridors (max dimension \(<2\) m \cite{standard}) and adopt a 5 m grid, sufficient for single-UAV operation.} The received sensing power threshold is $\epsilon_1 = -75 $ dBm, and the SINR threshold is $\epsilon_2 = 3$ dB.

We consider the following two benchmark schemes for comparison:
\begin{itemize}
  \setlength{\itemsep}{0pt} 
  \setlength{\parskip}{0pt}
  \setlength{\parsep}{0pt}
  \setlength{\topsep}{0pt} 
    \item \textbf{A*-based sequential planning}: We first compute the shortest path via A* \cite{A_star} and then, conditioning on this path, solve problem (P3.2) to obtain the BS deployment subject to the communication and sensing constraints.
    \item \textbf{Random BS deployment}: We randomly deploy BSs at the \(K\) candidate sites with the number of deployed BSs increasing from 1 to \(K\), and solve problem (P3.1) with given \(\boldsymbol{\delta}\) until a feasible shortest path solution is obtained.
\end{itemize}

Fig. 3 illustrates the jointly optimized coarse‐layer air corridor, BS deployment, and the received sensing power. It is observed that the cells traversed by the corridor satisfy both thresholds, and the length is minimized. On a $10\times10$ grid, our method yields a shortest path traversing 19 cells (50 m per cell) with 8 deployed BSs. Furthermore, Fig. 4 presents the fine‐layer corridors and BS deployment after iterative optimizations, with each fine cell according to the SINR. On a $100\times100$ grid, our method yields a shortest path traversing 199 cells (5 m per cell) with 7 deployed BSs.
\vspace{-4pt}
\begin{figure}[htbp]
  \centering
  \begin{minipage}[b]{0.48\linewidth}
    \includegraphics[width=\linewidth]{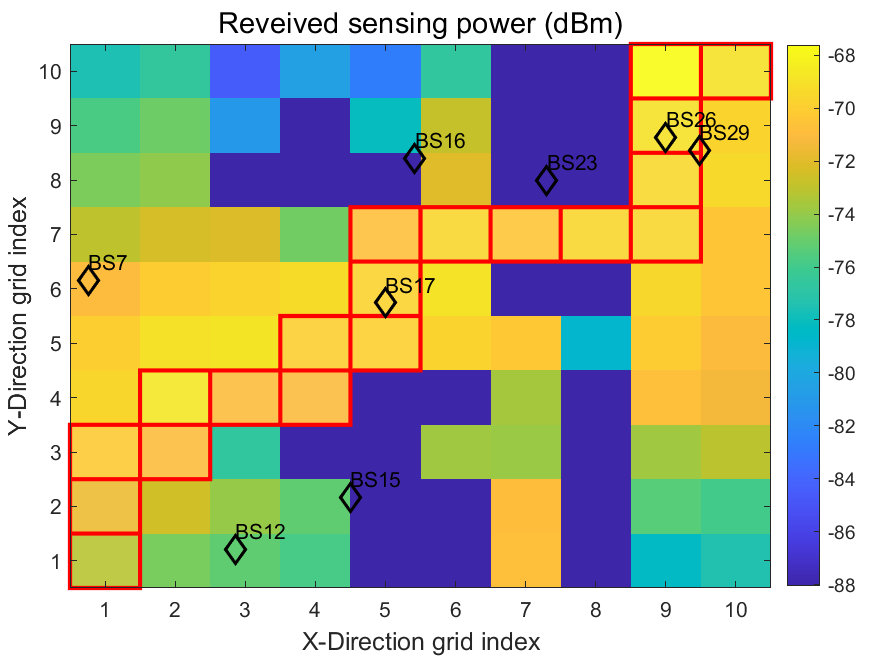}
    \captionof{figure}{Coarse‐layer air corridor overlaid on the sensing power map.}
    \label{fig:random}
  \end{minipage}\hfill
  \begin{minipage}[b]{0.48\linewidth}
    \includegraphics[width=\linewidth]{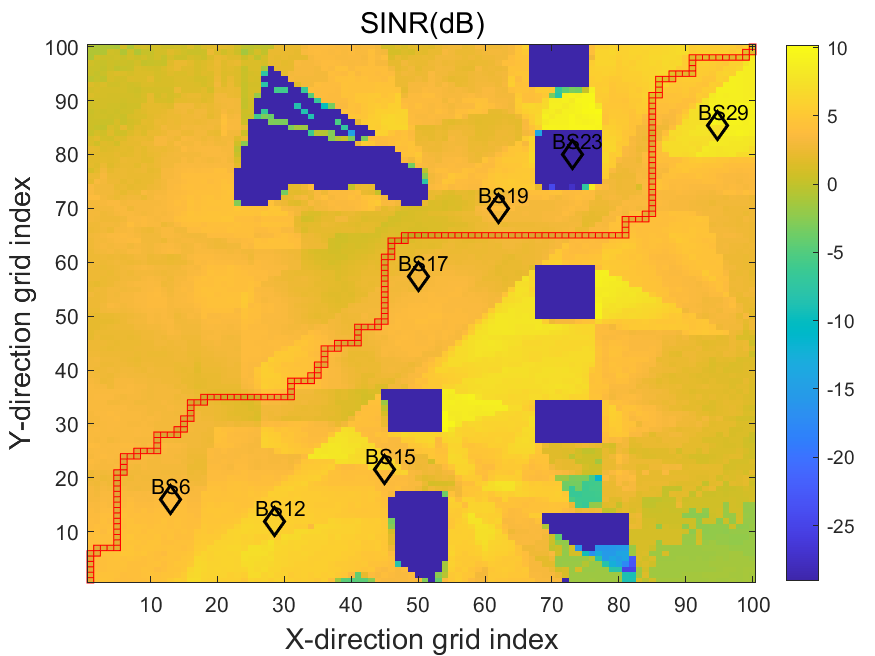}
    \captionof{figure}{Fine‐layer air corridor overlaid on the SINR map.}
    \label{fig:astar}
  \end{minipage}
\end{figure}

Fig. 5 shows the number of deployed BSs versus the minimum received sensing power threshold $\epsilon_1$ (with $\epsilon_2=3$ dB), and Fig. 6 shows the length versus the minimum SINR threshold $\epsilon_2$ (with $\epsilon_1=-75$ dBm) for our method and two benchmarks. The results for the two benchmarks are averaged over 100 random realizations. It is observed in Fig. 5 that the number of deployed BSs increases with \(\epsilon_1\) for all three schemes, whereas our approach consistently requires the fewest BSs. It is observed in Fig. 6 that the length of both our method and the random deployment baseline increases with $\epsilon_2$, whereas that of the A* scheme remains unchanged. Nevertheless, our approach consistently identifies a feasible path and yields the shortest length among all three schemes. By contrast, the A* scheme fails to find any feasible path once $\epsilon_2$ exceeds 3.2 dB, and the random BS deployment baseline fails beyond 3.8 dB, since under stringent SINR thresholds and limited trial budget, these methods rarely identify a path satisfying all constraints.
\begin{figure}[htbp]
  \centering
  \begin{minipage}[b]{0.48\linewidth}
    \includegraphics[width=\linewidth]{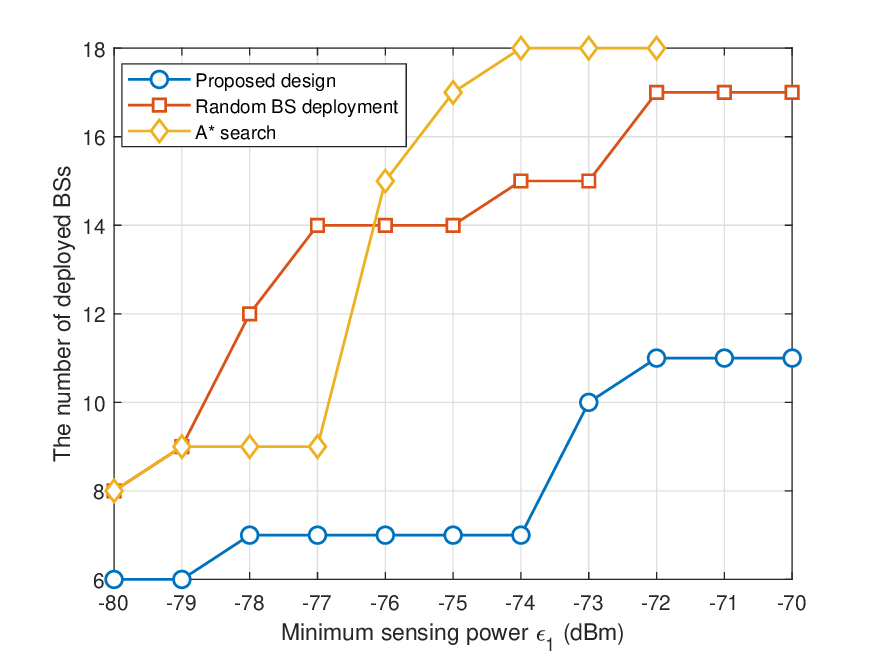}
    \captionof{figure}{The number of deployed BSs versus $\epsilon_1$.}
    \label{cost1}
  \end{minipage}\hfill
  \begin{minipage}[b]{0.48\linewidth}
    \includegraphics[width=\linewidth]{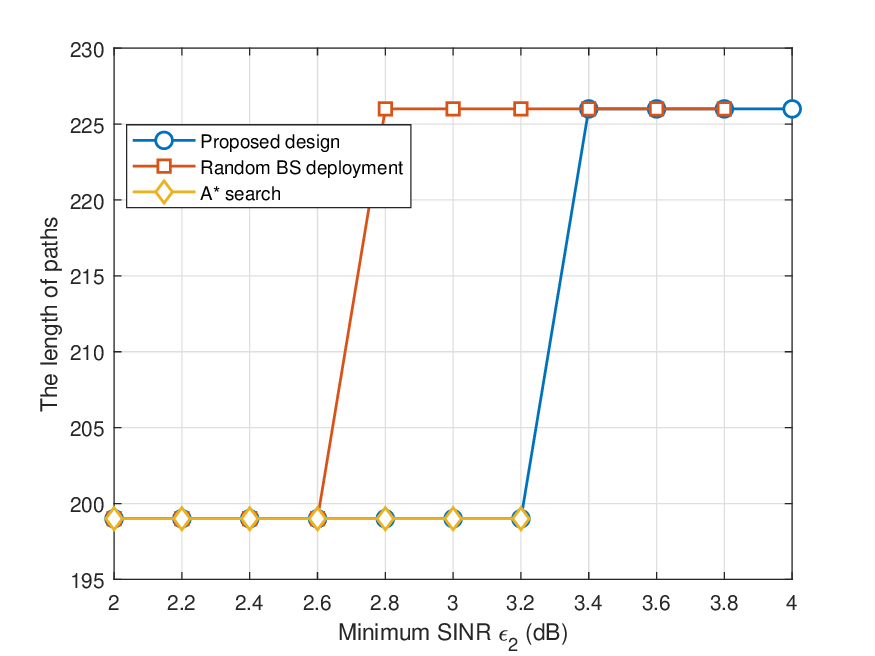}
    \captionof{figure}{The length of air corridor paths versus $\epsilon_2$.}
    \label{cost2}
  \end{minipage}
\end{figure}

\section{Conclusion}

In this letter, we investigated the joint air corridor planning and BS deployment for low-altitude ISAC networks, with the objective of minimizing the network cost under both communication and sensing constraints. By leveraging CKMs, we enabled offline evaluation of each candidate BS’s performance without the need for real-time measurements. We proposed a coarse-to-fine framework with an AO algorithm to efficiently solve the formulated large-scale, nonconvex integer program. Simulation results show that the proposed method significantly reduces length and deployment cost compared to benchmark schemes, thereby demonstrating its effectiveness for cost-efficient planning of low-altitude ISAC networks.

\vspace{-0.1\baselineskip}

%\fi
\iffalse
\begin{align}
(\text{P1}):\quad&\min_{\substack{\mathbf{A},\boldsymbol{\delta}}}\quad  \alpha_1 \mathbf{1}^{\top}\mathbf{A}\mathbf{1} + \alpha_2 \sum\nolimits_{k=1}^K \delta_k, \notag\\
    \text{s.t.}\;\;
    &\sum\nolimits_{k=1}^{K} \delta_k p_{k,i,j} \geq \epsilon_1 A_{i,j}, \ \forall i,j\in\mathcal{N},\tag{a}\\
    &\sum\nolimits_{k=1}^{K} \delta_k \mathcal{L}_{k,i,j} \geq 3 A_{i,j}, 
    \ \forall i,j\in\mathcal{N},\tag{b}\\
    &\max\nolimits_{k\in\mathcal{K}} \underline{\gamma}_{k,i,j}(\boldsymbol{\delta})\geq \epsilon_2 A_{i,j}, 
    \ \forall i,j\in\mathcal{N},\tag{c}\\
    &\delta_k \in \{0, 1\}, A_{i,j} \in \{0, 1\}, 
    \ \forall k\in\mathcal{K}, i,j\in\mathcal{N},\tag{d}\\
    &2 A_{i,j}\leq\sum\nolimits_{(i',j')\in \mathcal N_4(i,j)}A_{i',j'} \leq2,{e}\\
    &\sum\nolimits_{i=1}^{N} A_{i,j} \ge 1,\forall j\in\mathcal N,\sum\nolimits_{j=1}^{N} A_{i,j} \ge 1,\forall i\in\mathcal N,\tag{f} \\
    &A_{1,1}=1,\quad A_{N,N}=1.\tag{g}
\end{align}
\fi
%\vspace{12pt}


\begin{thebibliography}{99}
\bibitem{mobility}
A.~Bauranov and J.~Rakas, ``Designing airspace for urban air mobility: 
A review of concepts and approaches,'' \emph{Prog. Aerosp. Sci.}, vol.~125, p.~100726, 2021.
\bibitem{Integrated_Sensing_and_Communications}
F.~Liu, Y.~Cui, C.~Masouros, J.~Xu, T.~X.~Han, Y.~C.~Eldar, and S.~Buzzi,
``Integrated sensing and communications: Toward dual-functional wireless networks for 6G and beyond,''
\emph{IEEE J. Sel. Areas Commun.}, vol.~40, no.~6, pp.~1728--1767, Jun. 2022.
\bibitem{aircorridor}
S.~I.~Muna, S.~Mukherjee, K.~Namuduri, M.~Compere, M.~I.~Akbas, P.~Molnár, and R.~Subramanian,
``Air corridors: Concept, design, simulation, and rules of engagement,'' \emph{Sensors}, vol.~21, no.~22, Art. no. 7536, 2021.
\bibitem{coordinate}
L.~Yu, Z.~Li, N.~Ansari, and X.~Sun, 
``Hybrid transformer based multi-agent reinforcement learning for multiple unpiloted aerial vehicle coordination in air corridors,'' 
\emph{IEEE Trans. Mobile Comput.}, 
vol.~24, no.~6, pp.~5482--5495, Jun. 2025.
\bibitem{overview1}
Y.~Song, Y.~Zeng, Y.~Yang, Z.~Ren, G.~Cheng, X.~Xu, J.~Xu, S.~Jin, and R.~Zhang,
``An overview of cellular ISAC for low-altitude UAV: New opportunities and challenges,''
\emph{IEEE Commun. Mag.}, pp.~1--8, 2025.
\bibitem{overview2}
Q.~Wu, J.~Xu, Y.~Zeng, D.~W.~K.~Ng, N.~Al-Dhahir, R.~Schober, and A.~L.~Swindlehurst,
``A comprehensive overview on 5G-and-beyond networks with UAVs: From communications to sensing and intelligence,''
\emph{IEEE J. Sel. Areas Commun.}, vol.~39, no.~10, pp.~2912--2945, Oct.~2021.
\bibitem{Cooperative_ISAC_Networks}
G.~Cheng, X.~Song, Z.~Lyu, and J.~Xu, ``Networked ISAC for low-altitude economy: Coordinated transmit beamforming and UAV trajectory design,'' \emph{IEEE Trans. Commun.}, vol.~73, no.~8, pp.~5832--5847, Aug. 2025.
\bibitem{b5}
Y.~Zeng, J.~Chen, J.~Xu, D.~Wu, X.~Xu, S.~Jin, X.~Gao, D.~Gesbert, S.~Cui, and R.~Zhang, 
``A tutorial on environment-aware communications via channel knowledge map for 6G,'' 
\emph{IEEE Commun. Surveys Tuts.}, 
vol.~26, no.~3, pp.~1478--1519, 3rd Quart., 2024.
\bibitem{CKM1}
B. Zhang and J. Chen, \textquotedblleft Constructing Radio Maps for UAV Communications via Dynamic Resolution Virtual Obstacle Maps,\textquotedblright{} in \emph{Proc. 2020 IEEE 21st Int. Workshop Signal Process. Adv. Wireless Commun. (SPAWC)}, Atlanta, GA, USA, May 2020, pp.~1--5.
\bibitem{CKM2}
W. Liu and J. Chen, \textquotedblleft UAV-Aided Radio Map Construction Exploiting Environment Semantics,\textquotedblright{} \textit{IEEE Trans. Wireless Commun.}, vol. 22, no. 9, pp. 6341-6355, Sep. 2023.
\bibitem{b4}
Y.~Zeng and X.~Xu, 
``Toward environment-aware 6G communications via channel knowledge map,'' 
\emph{IEEE Wireless Commun.}, 
vol.~28, no.~3, pp.~84--91, Jun. 2021.
\bibitem{CKM}
H.~Li, P.~Li, G.~Cheng, J.~Xu, J.~Chen, and Y.~Zeng,
``Channel knowledge map (CKM)-assisted multi-UAV wireless network: CKM construction and UAV placement,''
\emph{J. Commun. Inf. Netw.}, vol.~8, no.~3, pp.~256--270, Sep.~2023.
\bibitem{sionna}
J.~Hoydis, F.~A.~Ait~Aoudia, S.~Cammerer, M.~Nimier-David, N.~Binder, G.~Marcus, and A.~Keller,
``Sionna RT: Differentiable Ray Tracing for Radio Propagation Modeling,''
in \emph{Proc. 2023 IEEE Globecom Workshops (GC Wkshps)}, 2023, pp.~317--321.
\bibitem{radar}
M.~A.~Richards, J.~A.~Scheer, and W.~A.~Holm,
\emph{Principles of Modern Radar: Basic Principles}.
Ann Arbor, MI, USA: IET Digital Library, 2010.
\bibitem{loc}
S. Motie, H. Zayyani, M. Salman, and M. Bekrani, ``Self UAV localization using multiple base stations based on TDoA measurements,'' 
\emph{IEEE Wireless Commun. Lett.}, vol. 13, no. 9, pp. 2432–2436, Jun. 2024.
\bibitem{bigM}
M.~S.~Bazaraa, J.~J.~Jarvis, and H.~D.~Sherali, 
\emph{Linear Programming and Network Flows}.
New York, NY: Wiley, 1990.
\bibitem{four}
G.~P.~McCormick, 
``Computability of global solutions to factorable nonconvex programs: Part I—Convex underestimating problems,'' 
\emph{Math. Program.}, vol.~10, no.~1, pp.~147--175, 1976.
\bibitem{convex}
S. Boyd and L. Vandenberghe, \emph{Convex Optimization}. Cambridge, U.K.:
Cambridge Univ. Press, 2004.
\bibitem{standard}
\emph{IEEE Standard for a Framework for Structuring Low-Altitude Airspace for Unmanned Aerial Vehicle (UAV) Operations}, Standard 1939.1-2021, 2021, pp. 1–94.
\bibitem{A_star}
P.~E.~Hart, N.~J.~Nilsson, and B.~Raphael,
``A Formal Basis for the Heuristic Determination of Minimum Cost Paths,''
\emph{IEEE Trans. Syst. Sci. Cybern.}, vol.~4, no.~2, pp.~100--107, 1968.

\end{thebibliography}
\end{document}